\documentclass[12pt, draftclsnofoot, onecolumn]{IEEEtran}
\usepackage{cite}
\usepackage{graphicx,amsmath,amssymb}
\usepackage{subfigure}
\usepackage{citesort}
\usepackage{fancyhdr}
\usepackage{mdwmath}
\usepackage{mdwtab}
\usepackage{balance}
\usepackage{xcolor}
\usepackage{bm}

\usepackage{algorithm}
\usepackage{algorithmic}
\usepackage{multirow}
\usepackage{flafter}
\usepackage{comment}

\usepackage{amsthm}

\usepackage{float}
\floatstyle{plaintop}
\restylefloat{table}

\newtheorem{remark}{Remark}
\newtheorem{theorem}{Theorem}
\newtheorem{lemma}{Lemma}

\newtheorem{corollary}{\textbf{Corollary}}

\begin{document}

%\title{Fairness of User Clustering in MIMO Non-orthogonal Multiple Access Systems}
%\author{Author 1,  Author 2, Author 3, and Author 4}
%%\markboth{IEEE Transactions on \LaTeX\ }
%%{Hayes}

\title{Simultaneously Transmitting and Reflecting (STAR)-RISs: Are They Applicable to Dual-Sided Incidence?}

%
% {\author{ Yuanwei\ Liu,  Maged\ Elkashlan, Zhiguo\ Ding, and George\ K. Karagiannidis}

\author{Jiaqi\ Xu, Xidong\ Mu, Joey Tianyi Zhou, and Yuanwei\ Liu.

\thanks{J. Xu, X. Mu, and Y. Liu are with the School of Electronic Engineering and Computer Science, Queen Mary University of London, London E1 4NS, UK. (email:\{jiaqi.xu, x.mu, yuanwei.liu\}@qmul.ac.uk).}
\thanks{J.T. Zhou is with A*STAR Centre for Frontier AI Research (CFAR), 138632, Singapore. (e-mail: joey\_zhou@ihpc.a-star.edu.sg).}
%\thanks{R. Schober is with the Institute for Digital Communications, Friedrich-Alexander-University Erlangen-N\"{u}rnberg (FAU), Germany (e-mail: robert.schober@fau.de).}

}
\maketitle
%\thispagestyle{fancyplain}
%\pagestyle{fancy}
\begin{abstract}
A hardware model and a signal model are proposed for \textit{dual-sided}
simultaneously transmitting and reflecting reconfigurable intelligent surfaces (STAR-RISs), where the signal simultaneously incident on both sides of the surface.
Based on the proposed hardware model, signal models for dual-sided STAR-RISs are developed. 
For elements with scalar surface impedance, it is proved that their transmission and reflection coefficients on both sides are \textit{identical}.
Based on the obtained symmetrical dual-sided STAR model, a STAR-RIS-aided two-user uplink communication system is investigated for both non-orthogonal multiple access (NOMA) and orthogonal multiple access (OMA) schemes. 
Analytical results for the outage probabilities for users are derived in the high transmit signal-to-noise ratio (SNR) regime.
Numerical results demonstrate the performance gain of NOMA over OMA and reveal that the outage probability error floor can be lowered by adjusting the ratio between the amplitudes of transmission and reflection signals.
\end{abstract}


\section{Introduction}

To facilitate full-space smart radio environment, the novel concept of simultaneously transmitting and reflecting reconfigurable intelligent surfaces (STAR-RISs)~\cite{liu_star,xu_star,IOS,added,mu_star} has been recently proposed. Unlike the reflecting-only RISs or known as intelligent reflecting surfaces (IRSs), STAR-RISs are capable of supporting simultaneous transmission and reflection\footnote{Here, transmission refers to the physical process of signal penetrating the surface.}. Thus, the wireless signal can propagate to both sides of the surface while its amplitude and phase are beneficially controlled during the transmission and reflection.
In~\cite{xu_star}, the authors proposed a hardware model and two channel models for employing STAR-RISs to assist downlink communications.
%
%As a further advance, the authors of \cite{mu_star} proposed three tailored practical operating protocols and studied the corresponding beamforming design for STAR-RIS assisted multi-antenna multi-user downlink communications.
%
Although the benefits of STAR-RISs have been demonstrated, existing works mainly focused on the STAR-RIS-aided downlink transmission. Recall the fact that the full-space coverage can be provided by STAR-RISs. However, it is not yet clear how wireless signals can be manipulated when the signals incident on both sides of the surface simultaneously. This is practically relevant to the uplink communication scenario, where multiple users surrounding the STAR-RIS have to upload their information to the base station (BS).

\textcolor{black}{
Against the above background, this work compliments the downlink signal model we proposed in~\cite{xu_star} and provides the hardware and signal models for
STAR-RIS under dual-sided signal incidence. 
Exploiting electromagnetic (EM) theory, we confirm that STAR-RISs are able to simultaneously reconfigure the propagation of signals incident on different sides. In this letter, we use the term \textit{dual-sided} STAR-RISs to distinguish this dual-side working mode with the single-sided signal incident working mode.}
Specifically, for the case where the surface impedance of each element can be characterized with scalars, we prove that the transmission and reflection coefficients on both sides are identical, meaning that each STAR-RIS element is symmetrical. Based on the proposed models, a STAR-RIS-aided two-user uplink communication system is investigated with non-orthogonal multiple access (NOMA) and orthogonal multiple access (OMA). 
\textcolor{black}{Due to the ability to beneficially adjust the power ratio of the uploaded signals between users on transmission and reflection sides of the STAR-RIS, STAR-RIS-aided uplink NOMA is capable of achieving higher gain over conventional reflecting-only RIS-aided NOMA.}
We provide the new channel statistics for the double-Rician cascaded channels. Moreover, closed-form expressions are given for the outage probabilities of uplink users within the high-SNR regime.

\section{EM-Based Modeling for Dual-Sided STAR-RISs}
In prior studies, two complex coefficients, namely, the transmission and reflection (T\&R) coefficients, are introduced to characterize the \textit{STAR} feature of each element~\cite{xu_star}.
However, these two coefficients are derived by assuming that the signals only incident from one side of the surface and there is no signal incident from the other side. The signal model is generally unknown for the case where there are wireless signals simultaneously incident on both sides of the STAR-RIS, namely \emph{dual-sided incidence}. To fill this knowledge gap, in this section, 
we provide a signal model for \textit{dual-sided} STAR-RISs. Since the `STAR' process occurs on each side, we introduce two pairs of coefficients in the following general signal model. 
\begin{figure}[h!]
    \begin{center}
        \includegraphics[scale=0.4]{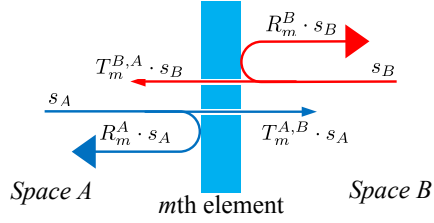}
        \caption{A general signal model for dual-sided STAR-RISs}
        \label{randt}
    \end{center}
\end{figure}

\begin{figure}[h!]
    \begin{center}
        \includegraphics[scale=0.3]{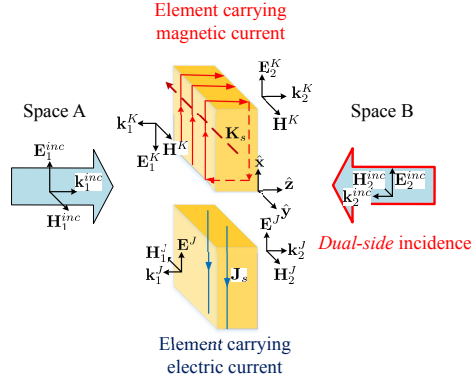}
        \caption{Schematic of the STAR-RIS element under EM wave incidence.}
        \label{ele}
    \end{center}
\end{figure}

\begin{theorem}
As illustrated in Fig.~\ref{randt}, for vertically polarized wireless signals, the signals radiating from the $m$th element towards \textit{Space A} ($s'_A$) and towards \textit{Space B} ($s'_B$) \textcolor{black}{are linear superposition of the incident signals ($s_{A/B}$):}
\textcolor{black}{
\begin{align}\label{signal_model}
    s'_A = \sum_{\theta \in \Theta_A}R^A_m(\theta) \cdot s_A(\theta) + \sum_{\theta \in \Theta_B }T^{B,A}_m(\theta) \cdot s_B(\theta), \\
    s'_B = \sum_{\theta \in \Theta_A}T^{A,B}_m(\theta) \cdot s_A(\theta) + \sum_{\theta \in \Theta_B}R^B_m(\theta) \cdot s_B(\theta), 
\end{align}
}where $T^{A,B}_m(\theta)$ and $T^{B,A}_m(\theta)$ are the transmission coefficients from \textit{Space A/B} to \textit{Space B/A}, $R^{A/B}_m(\theta)$ is the reflection coefficient for \textit{Space A/B}, subscript $m$ denote the $m$th STAR element, \textcolor{black}{$\theta$ is the incidence angle of the signal, and $\Theta_A$ and $\Theta_B$ are the set of angles of the incident signals from Space A and Space B}, respectively.
\begin{proof}
The key for this proof is to demonstrate that the received signals are the linear superposition of the incident signals from different directions. According to \cite{estakhri2016wave}, for STAR-RIS that is electrically thin, the EM response can be equivalently described using surface electric and magnetic currents $\mathbf{J}_s$ and $\mathbf{K}_s$. 
\textcolor{black}{
Note that the direction of $\mathbf{K}_s$ is determined through the curl right-hand rule, which is illustrated in Fig.~\ref{ele}.}
Considering passive STAR-RIS elements, these currents are excited by the total EM field $\mathbf{E}_{tot},\mathbf{H}_{tot}$, and scaled by the electric admittance ($\overline{\overline{\mathbf{Y}}}_e$) and magnetic impedance ($\overline{\overline{\mathbf{Z}}}_m$) tensors at each element, i.e., $\mathbf{J}_s = \overline{\overline{\mathbf{Y}}}_e \cdot \mathbf{E}_{tot}$ and $\mathbf{K}_s = \overline{\overline{ \mathbf{Z}}}_m \cdot \mathbf{H}_{tot}$. As illustrated in Fig.~\ref{ele}, the electric current density $\mathbf{J}_s$ generates fields $\mathbf{E}^J$, $\mathbf{H}^J_1$, and $\mathbf{H}^J_2$ while the magnetic current density $\mathbf{K}_s$ generates fields $\mathbf{E}^K_1$, $\mathbf{E}^K_2$, and $\mathbf{H}^K$. The magnitude of the radiating field, $\mathbf{E}^J$ and $\mathbf{E}^K$, are proportional to the strength of $\mathbf{J}_s$ and $\mathbf{K}_s$. Thus, \textcolor{black}{for signals with the same polarization,} the received field can be expressed as the linear combination of the incident fields $\mathbf{E}^{inc}_1$ and $\mathbf{E}^{inc}_2$. Finally, by using the fact that the signal $s_{A/B}$ is the proportional to the magnitude of the corresponding electric field, i.e., $s_{A/B} = ||\textbf{E}^{inc}_{A/B}||_2$, the theorem is derived.
\end{proof}
\end{theorem}

\begin{figure*}[t!]
\centering
\subfigure[Purely transmitting configuration (connecting \textit{Space A/B})]{\label{sysa}
\includegraphics[width= 2in]{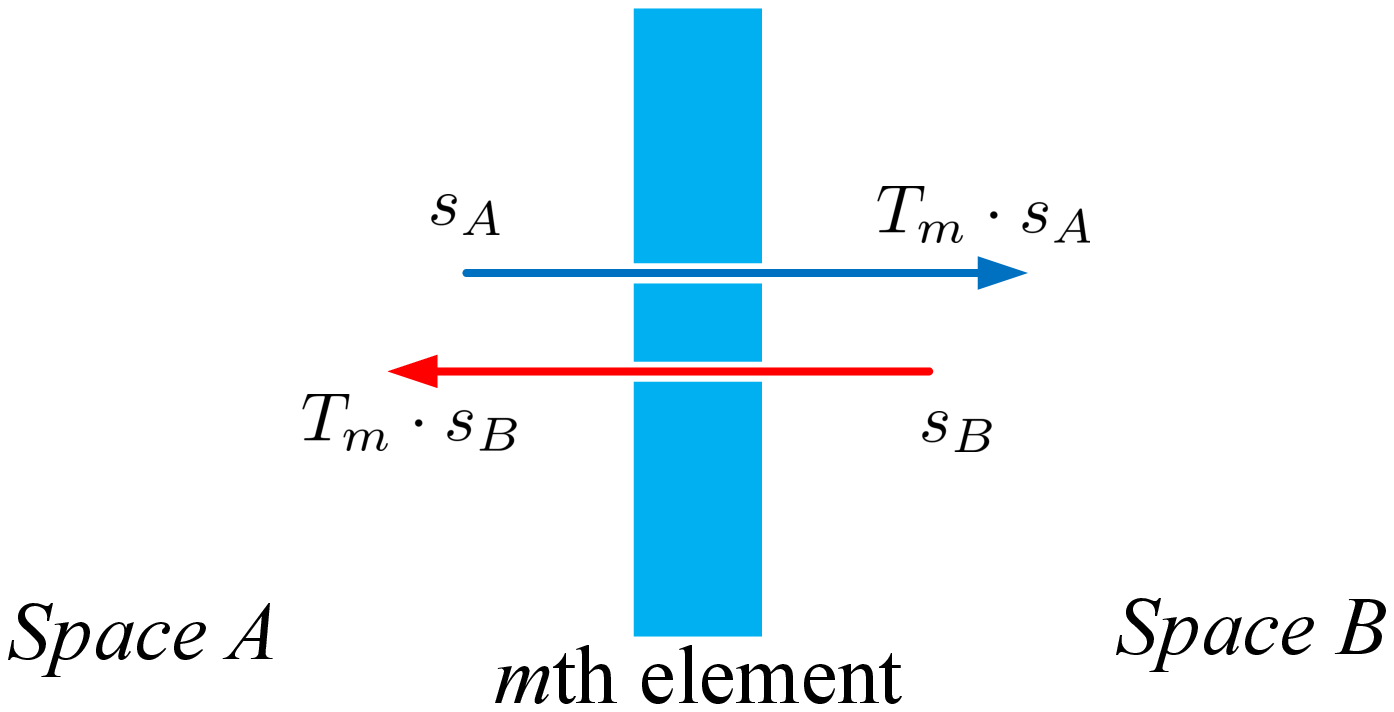}}
\subfigure[Purely reflecting configuration (separating \textit{Space A/B})]{\label{sysb}
\includegraphics[width= 2in]{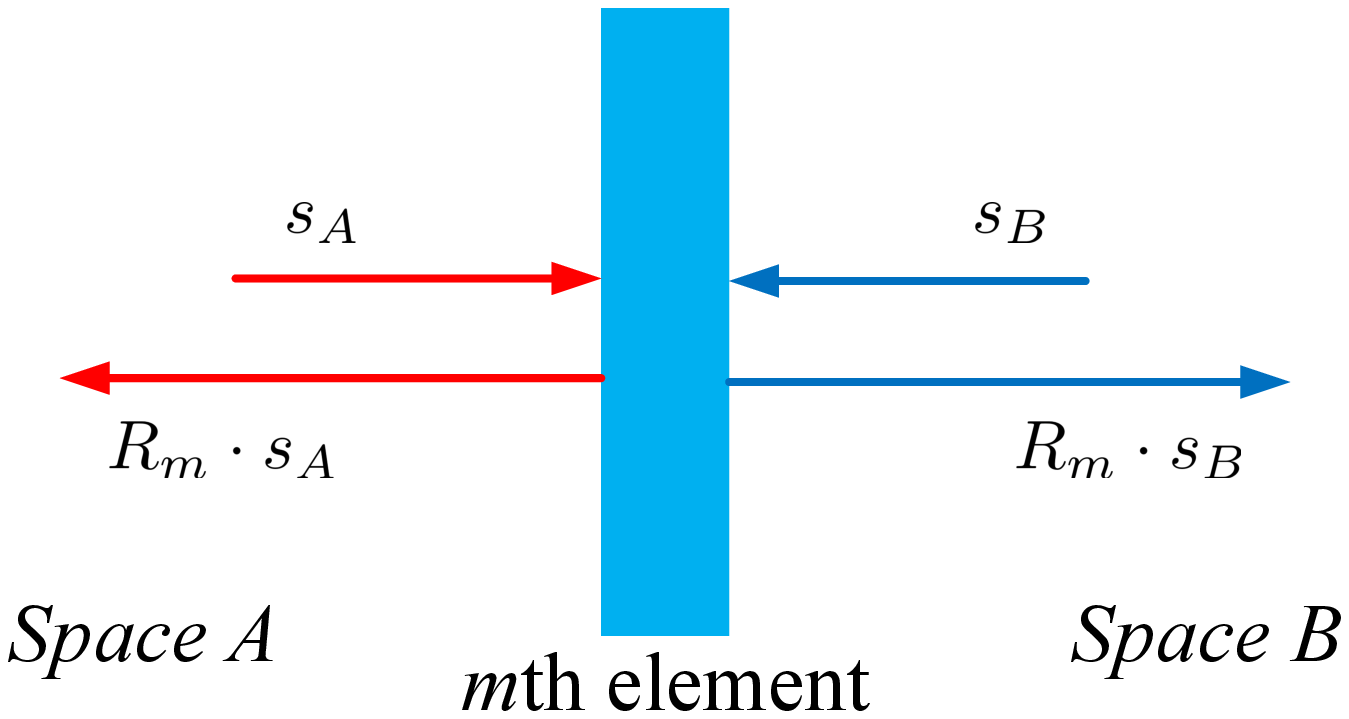}}
\caption{Signal model for purely transmitting and purely reflecting.}\label{nice0}
\end{figure*}

\subsection{A Hardware Model for Symmetrical Dual-Sided STAR-RISs}
For a general scenario, the T\&R coefficients for the two sides can be different, i.e., $T^{A,B}_m \neq T^{B,A}_m$ and $R^A_m\neq R^B_m$. However, for the case where the surface impedances can be characterized with scalars~\cite{estakhri2016wave}, the elements have symmetrical EM response for the two directions, \textcolor{black}{i.e., $T^{A,B}_m = T^{B,A}_m$ and $R^A_m = R^B_m$.}
Following the method as adopted in~\cite{sci_rep,xu_correlated}, we separately study the radiation of electric and magnetic currents\footnote{Here, the magnetic current refers to the equivalent magnetic current generated by the vortex currents~\cite{old}.}. The schematic illustration of the STAR-RIS under EM incidence on both sides is presented in Fig.~\ref{ele}. 
In \textit{Space A/B}, the total electrical radiation comes from the contributions of these induced electric fields and the overall incident fields $\mathbf{E}^{inc}_1$ and $\mathbf{E}^{inc}_2$.
The strengths of the surface electric and magnetic currents are proportional to the sum of the incident and radiated fields at the STAR-RIS~\cite{sci_rep}.
Thus, the density of the electric and magnetic currents are respectively given by:
\begin{align}\label{js}
    \mathbf{J}_s &= (\mathbf{E}^{inc}_1+\mathbf{E}^{inc}_2+\mathbf{E}^J)/Z_e,\\ \label{ks}
    \mathbf{K}_s &=(\mathbf{H}^{inc}_1-\mathbf{H}^{inc}_2+\mathbf{H}^K)Z_m,
\end{align}
where $Z_e$ and $Z_m$ are the scalar electric and magnetic impedance of the $m$th element, respectively.
To solve the generated electric fields $\mathbf{E}^{J}$ and $\mathbf{E}^{K}$, we consider the following boundary conditions.
The boundary conditions~\cite{rothwell2018electromagnetics} of EM field at the surface can be expressed as $\hat{z}\times (\mathbf{H}^J_2-\mathbf{H}^J_1)\hat{y} = \mathbf{J}_s\hat{x}$ and \textcolor{black}{ $\hat{z}\times(\mathbf{E}^K_1 - \mathbf{E}^K_2)\hat{x}=\mathbf{K}_s\hat{y}$}. Thus, we can obtain the following:
\begin{align}
&\mathbf{E}^{J} = -\frac{\eta(\mathbf{E}^{inc}_1+\mathbf{E}^{inc}_2)}{2Z_e+\eta},\\
&\mathbf{E}^{K}_1 = - \mathbf{E}^{K}_2 = \frac{Z_m(\mathbf{E}^{inc}_1-\mathbf{E}^{inc}_2)}{Z_m+2\eta},
\end{align}
where $\eta$ is the free space wave impedance.

For a receiver located in \textit{Space A/B}, the received signal strength is proportional to the square of the overall received electric field. As illustrated in Fig.~\ref{ele}, for \textit{Space A}, we collect all the wave vectors going towards the left, i.e., $\mathbf{k}^{inc}_2$, $\mathbf{k}^{J}_1$, and $\mathbf{k}^{K}_1$. Thus, the overall electric field propagating towards \textit{Space A/B} through the $m$th STAR element can be calculated as follows:
\begin{align}
\begin{split}
    \mathbf{E}^{rec}_{A} &= \mathbf{E}^J + \mathbf{E}^K_{1} + \mathbf{E}^{inc}_{2} =
    \underbrace{\big( \frac{Z_m}{Z_m+2\eta}-\frac{\eta}{2Z_e+\eta}\big)}_{R_m}\mathbf{E}^{inc}_{1} 
    \\&+\underbrace{\big( \frac{2Z_e}{2Z_e+\eta} - \frac{Z_m}{Z_m+2\eta}\big)}_{T_m}\mathbf{E}^{inc}_{2}. \label{ea}
    \end{split}
\end{align}
Similarly, the overall electric field propagating towards \textit{Space B} is as follows:
\begin{align}
\begin{split}
    \mathbf{E}^{rec}_B &= \mathbf{E}^J + \mathbf{E}^K_2 + \mathbf{E}^{inc}_1 =
    \underbrace{\Big(\frac{2Z_e}{2Z_e+\eta} - \frac{Z_m}{Z_m+2\eta} \Big)}_{T_m}\mathbf{E}^{inc}_1 
    \\&+\underbrace{\Big( \frac{Z_m}{Z_m+2\eta}-\frac{\eta}{2Z_e+\eta}\Big)}_{R_m}\mathbf{E}^{inc}_2.
    \end{split}\label{eb}
\end{align}

Next, exploiting \eqref{ea} and \eqref{eb}, the signal model for dual-sided STAR-RIS can be established.

%
%%%%%
\subsection{Signal Modeling for Symmetrical Dual-Sided STAR-RIS Elements}
According to \eqref{ea} and \eqref{eb}, we have the following theorem.

\begin{theorem}
For symmetrical dual-sided STAR-RISs, the signals radiating from the $m$th element towards \textit{Space A} ($s'_A$) and towards \textit{Space B} ($s'_B$) are as follows:
\begin{align}\label{spa}
    s'_A = R_m \cdot s_A + T_m \cdot s_B, \\ \label{spb}
    s'_B = T_m \cdot s_A + R_m \cdot s_B, 
\end{align}
where
\begin{align} \label{ttm}
    T_m = \frac{2Z_e}{2Z_e+\eta} - \frac{Z_m}{Z_m+2\eta},\\ \label{rrm}
    R_m = \frac{Z_m}{Z_m+2\eta}-\frac{\eta}{2Z_e+\eta} 
\end{align}
are the T\&R coefficients for each STAR-RIS element.
\begin{proof}
\textcolor{black}{
The proof of this theorem is based on the boundary conditions and the density of the electric and magnetic currents given in \eqref{js} and \eqref{ks}. Exploiting these characteristics, the field in both \textit{Space A} and \textit{Space B} were given in \eqref{ea} and \eqref{eb}.
Then, by using the fact that the received signal $s'_{A/B}$ is the proportional to the magnitude of the received electric field, i.e., $s'_{A/B} = ||\textbf{E}^{rec}_{A/B}||_2$, \eqref{ea} and \eqref{eb} can be rewritten into the forms as in \eqref{spa} and \eqref{spb}, respectively.}
\end{proof}
\end{theorem}

\begin{comment}
\begin{corollary}
For the purely transmitting or purely reflecting scenario, the received signal in \textit{Space A} and \textit{Space B} are respectively given as follows:
\begin{equation}\label{c1eq1}
    s'_A = T_ms_B,\  s'_B = T_ms_A,
\end{equation}
for purely transmitting where $T_m = \frac{2Z_e}{2Z_e+\eta} - \frac{Z_m}{Z_m+2\eta}$ and
\begin{equation}\label{c1eq2}
    s'_A = R_m s_A,\  s'_B = R_m s_B,
\end{equation}
for purely reflecting, where $R_m = \frac{Z_m}{Z_m+2\eta}-\frac{\eta}{2Z_e+\eta}$.

\begin{proof}
For the purely transmitting scenario, the surface impedance of each element need to satisfy $\frac{Z_m}{Z_m+2\eta}=\frac{\eta}{2Z_e+\eta}$~\cite{estakhri2016wave}. By substituting this equation into \eqref{ea} and \eqref{eb}, \eqref{c1eq1} can be derived. Similarly \eqref{c1eq2} can be derived by plugging $\frac{2Z_e}{2Z_e+\eta} = \frac{Z_m}{Z_m+2\eta}$ in \eqref{ea} and \eqref{eb}.
\end{proof}

\end{corollary}

\end{comment}

\begin{figure}[b!]
    \begin{center}
        \includegraphics[scale=0.25]{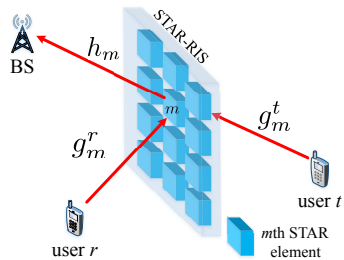}
        \caption{System model for the dual-sided STAR-RIS-aided uplink communication system.}
        \label{sys_fig}
    \end{center}
\end{figure}

\section{Performance Evaluation for A Dual-Sided STAR-RIS Uplink System}
Based on the above signal model, we evaluate the performance of a dual-sided STAR-RIS-aided uplink communication system. \textcolor{black}{We consider the two-user case and the employment of two fundamental multiple access schemes, i.e., OMA and NOMA. 
}

\subsection{System Model}

We study the application scenario where a symmetrical \textit{dual-sided} STAR-RIS is deployed to assist the uplink communication. 
\textcolor{black}{
To reveal the fundamental performance limit and to obtain closed-form results, we consider a two-user setup.}
We term the user which uploads the information via the reflection link as user $r$ and the user which uploads the information via the transmission link as user $t$. The direct link between the users and BS are assumed to be blocked. 
\textcolor{black}{
The STAR-RIS has $M$ elements whose complex-valued T\&R coefficients are given according to \eqref{ttm} and \eqref{rrm}. For convenience, in the following, we rewrite these coefficients in terms of their amplitudes and phase-shift arguments, i.e., $T_m = \beta^T_m e^{j\phi^T_m}$ and $R_m = \beta^R_m e^{j\phi^R_m}$, $m \in \{1,2,\cdots, M \}$.
}
The links between the $m$th STAR-RIS element and the BS, $h_m$, is assumed to follow Rician distribution, i.e., $|h_m| \sim \mathcal{R}(K_h,\Omega_h)$, where $K_h$ and $\Omega_h$ are the shape and scale parameters of the Rician distribution. The links between the $m$th STAR-RIS element and the users, $g^t_m$ and $g^r_m$, also follow Rician distribution, i.e., $|g^\chi_m|\sim \mathcal{R}(K^\chi_g,\Omega^\chi_g)$, where the notation $\chi\in\{t,r\}$ is an indicator representing the corresponding values for user $t$ or user $r$. 
For the multiple access schemes, we first consider NOMA and then consider conventional OMA as a baseline.

%Specifically, we consider the performance upper bound of the NOMA scheme. As a baseline scheme, we also provide the performance of TDMA scheme, as well as the scenario where STAR-RIS adopts T\&R-group phase-shift configuration strategy~\cite{xu_correlated}.

\subsection{NOMA}\label{sec_noma}
In NOMA, the two users upload their information to the BS via the same time/frequency resource blocks. The channel conditions of users depend on the choice of $T_m$ and $R_m$. According to the proposed signal model, the signal received at BS can be expressed as follows:
\begin{equation}\label{y}
    y = \sum_{m=1}^M g^r_m  h_m \beta^R_m e^{j\phi^R_m}  \sqrt{p} s_r
     + \sum_{m=1}^M g^t_m  h_m \beta^T_m e^{j\phi^T_m}  \sqrt{p} s_t + n_0,
\end{equation}
where $s_r$ and $s_t$ are the message of user $r$ and user $t$, respectively, $p$ is the transmit power of the uplink users, and $n_0$ is the additive white Gaussian noise at the BS with variance $\sigma^2_0$. Withough loss of generality, we assume user $r$ is closer to the STAR-RIS and has a higher channel gain than user $t$.
At the BS, the signal of user $r$ is decoded first by treating the other signal as interfering noise, and the corresponding signal-to-interference-plus-noise ratio (SINR) is given by:
\begin{equation}\label{sinrk}
    \mathrm{SINR}_r = \frac{p \big|\sum_{m=1}^M g^r_m  h_m \beta^R_m e^{j\phi^R_m}\big|^2}{p \big| \sum_{m=1}^M g^t_m  h_m \beta^T_m e^{j\phi^T_m} \big|^2 + \sigma^2_0}. 
\end{equation}
\textcolor{black}{
Then, the signal of user $t$ can be decoded after carrying out successive interference cancellation (SIC). Consider imperfect SIC, only a portion of the signal from user $r$ can be successfully cancelled. Thus, the signal of user $t$ has the following SINR:
\begin{equation}\label{snr0}
    \mathrm{SINR}_t = \frac{p\big| \sum_{m=1}^M g^t_m  h_m \beta^T_m e^{j\phi^T_m} \big|^2
    }{\alpha_0\cdot p\big| \sum_{m=1}^M g^t_m  h_m \beta^T_m e^{j\phi^T_m} \big|^2 + \sigma^2_0},
\end{equation}
where $\alpha_0\in[0,1]$ is the error propagation factor which characterizes the quality of the SIC, i.e., $\alpha_0=0$ corresponds to perfect SIC and $\alpha_0=1$ corresponds to the case where no cancellation is performed.
}
\subsection{New Channel Statistics}
New channel statistics can be obtained for the considered STAR-RIS-aided uplink system. Since the overall channel distribution is highly sensitive on the choice of $T_m$ and $R_m$, we consider the performance achieved by configuring $\phi^T_m$ and $\phi^R_m$ according to the cophase condition~\cite{di2019smart}. Moreover, we assume that the amplitudes, $\beta^T_m$ and $\beta^R_m$, are uniform across the entire STAR-RIS, i.e., $\beta^{T/R}_m = \beta^{T/R}, \ \forall m\in \{1,2,\cdots, M\}$. According to \eqref{sinrk} and \eqref{snr0}, under cophasing phase shift, the overall uplink channel for the two users can be expressed as follows:
\begin{align}\label{Hr}
    H_{r/t} &= \sum_{m=1}^M\beta^{R/T}|h_m|\cdot |g^{r/t}_m|.
\end{align}
Note that $\phi^{R/T}_m$  are chosen to align the phases of each terms in the sum of \eqref{Hr}. Next, we study the distribution of $|H_r|$, $|H_t|$, $\mathrm{SINR}_r$, and $\mathrm{SNR}_t$.

\begin{lemma}
Each term in the summation of \eqref{Hr} is the product of two Rician variables. Let $X^r_m = |h_m||g^r_m|$ and $X^t_m = |h_m||g^t_m|$, their probability density functions (PDFs) are given as follows:
\begin{align}\label{pro_rician}
\begin{split}
    &f_{|X^\chi_m|}(x) = \frac{x}{\beta_h\beta^\chi_g} e^{-(K_h+K^\chi_g)}\\
    &\cdot\sum_{i=0}^\infty\sum_{l=0}^\infty\frac{\big(\frac{\alpha_h\sqrt{x}}{2\beta_h} \big)^{2i}\big(\frac{\alpha^\chi_g\sqrt{x}}{2\beta^\chi_g} \big)^{2l}}{i!l!\Gamma(i+1)\Gamma(l+1)}
     \big(\frac{\beta_h}{\beta^\chi_g}\big)^{\frac{i-l}{2}}K_{i-l}\big(\frac{x}{\sqrt{\beta_h\beta^\chi_g}}\big),
    \end{split}
\end{align}
where $\alpha^2_{h}=\frac{K_{h}\Omega_{h}}{K_{h}+1}$,$(\alpha^\chi_g)^2=\frac{K^\chi_g\Omega^\chi_g}{K^\chi_g+1}$, $\beta_{h}=\frac{\Omega_{h}}{2(K_{h}+1)}$
, $\beta^\chi_{g}=\frac{\Omega^\chi_g}{2(K^\chi_g+1)}$, $\Gamma(x)$ denotes the Gamma function, and $K_n$ is the modified Bessel function of the second kind.
\begin{proof}
The detailed proof for the PDF of the product of two Rician variables can be found in \cite{simon2002probability}.
\end{proof}
\end{lemma}

\subsection{Outage Probability}
In the considered system, the achievable rate of user $r$ is given by $R_r = \log_2(1+\mathrm{{SINR}_r})$.
Consider a fixed-rate transmission, the outage probability of user $r$ is given by:
\begin{equation}\label{pout_r}
    P^r_{out} = \mathrm{Pr}\{ \mathrm{SINR}_r < 2^{\Tilde{R}_r}-1 \}.
\end{equation}
For user $t$, the BS needs to subtract the decoded message from user $r$ first, then decode the remaining message, thus, the outage probability is given by:
\begin{equation}\label{pout_t}
    P^t_{out} = 1 - \mathrm{Pr}\{ \mathrm{SINR}_r > 2^{\Tilde{R}_r}-1, \mathrm{SNR}_t > 2^{\Tilde{R}_t}-1 \},
\end{equation}
where $\Tilde{R}_r$ and $\Tilde{R}_t$ are the target data rate for user $r$ and $t$, respectively.
However, the PDF of the SINR in \eqref{sinrk} is difficult to evaluate in closed-form expressions. To reveal more practical insights, we study the outage probabilities in the high transmit SNR regime\footnote{Here, the transmit SNR is $\gamma_{transmit} = p/\sigma^2_0$}.

\subsubsection{High Transmit SNR Regime}
For the high transmit SNR scenario, we assume that $p/\sigma^2_0 \gg 1$. Thus, the outage probabilities of user $r$ can be reduced to:
\begin{equation}\label{prout}
    P^r_{out} = \mathrm{Pr}\Big\{\frac{\big|\sum_{m=1}^M g^r_m  h_m \beta^R_m e^{j\phi^R_m}\big|^2}{\big| \sum_{m=1}^M g^t_m  h_m \beta^T_m e^{j\phi^T_m} \big|^2} < \gamma_r \Big\},
\end{equation}
where $\gamma_r = 2^{\Tilde{R}_r}-1$.
\textcolor{black}{Considering the impact of imperfect SIC, the outage probability of user $t$ is given as follows:
\begin{equation}\label{ptout}
    P^t_{out} = \begin{cases}
       \mathrm{Pr}\Big\{\frac{\big|\sum_{m=1}^M g^r_m  h_m \beta^R_m e^{j\phi^R_m}\big|^2}{\big| \sum_{m=1}^M g^t_m  h_m \beta^T_m e^{j\phi^T_m} \big|^2} \!<\! \gamma_r \Big\},\!&\text{if}\ \alpha_0^{-1} > 2^{\tilde{R}_t}\!-\!1 \\
      1,&\text{otherwise}.
    \end{cases}
\end{equation}
}
In the following theorem, we give the closed-form analytical result for the outage probabilities.
\begin{theorem}\label{t3}
In the high transmit SNR regime, the outage probability is given by:
\begin{align}\label{P_high_snr}
    P_{out} = \Phi\left(\frac{\mu^tx_0 - \mu^r}{\sigma^t\sigma^r\xi(x_0)} \right),
\end{align}
where $\Phi(y) = \int_{u=-\infty}^y \frac{1}{\sqrt{2\pi}}e^{-u^2/2} du$ is the error function, $\xi(x) = \sqrt{\frac{x^2}{(\sigma^r)^2}-\frac{2\rho x}{\sigma^r\sigma^t}+\frac{1}{(\sigma^t)^2}}$, \textcolor{black}{$\rho = \mathrm{Cov}(|H_r|,|H_t|)/(\sigma^r\sigma^t)$ is the correlation coefficient between $|H_r|$ and $|H_t|$}, $x_0 = \sqrt{\gamma_r}\beta^T/\beta^R$ is the threshold for the outage probability, $\mu^\chi = \frac{1}{4}\sqrt{\frac{\pi\Omega_{h}}{(K_{h}+1)}}L_{1/2}(-K_{h})\sqrt{\frac{\pi\Omega^\chi_{g}}{(K^\chi_{g}+1)}}L_{1/2}(-K^\chi_{g})$ is the expectation value of $|h_m||g^\chi_m|$, and $(\sigma^\chi)^2 = \Omega_h\Omega^\chi_g - (\mu^\chi)^2$ is its variance.
\begin{proof}
According to the central limit theorem, both $H_r$ and $H_t$ follow the normal distribution when $M$ is sufficiently large.
\textcolor{black}{
In the high transmit SNR regime, the outage probabilities in both \eqref{prout} and the successful case of \eqref{ptout} are reduced to $P_{out} = \mathrm{Pr}\{|H_r|/|H_t| < \sqrt{\gamma_r}\}$. According to \eqref{Hr}, $|H_r|$ and $|H_t|$ have common terms and we denote their correlation coefficient as $\rho$. According to \cite{ratio}, the cumulative distribution function (CDF) of the ratio between two correlated normal distributions can be expressed in terms of the error function.}
\end{proof}
\end{theorem}

\begin{remark}
As shown in \eqref{P_high_snr}, the analytical outage probabilities do not further decrease with the increase of the uploading transmit power $p$, this indicates that the outage probabilities of both users approach the same error floor.
\end{remark}

\textcolor{black}{
\begin{remark}
In Theorem~\ref{t3}, we use a normal distribution ($f_{\text{Gaus}}(x)$) to approximate the PDF of $|H_r|$ which the exact distribution ($f_{\text{exact}}(x)$) is the sum of $M$ cascaded Rician channels. The approximation error decreases with the increase of $M$. In the following table, we calcualted the Kullback–Leibler divergence (relative entropy) from $f_{\text{exact}}(x)$ to $f_{\text{Gaus}}(x)$, i.e., $D_{\text{KL}} = \int_{x=-\infty}^\infty f_{\text{Gaus}}(x)\cdot \log\left( \frac{f_{\text{Gaus}}(x)}{f_{\text{exact}}(x)}\right)$. It can be observed that $D_{\text{KL}}$ quickly diminishes after $M\geq 20$.
\begin{table}[!h]
\centering
%\resizebox{\columnwidth}{!}{%
\begin{tabular}{|c|c|c|c|c|c|}
\hline
$M$ & 5      & 10     & 15       & 20                  & 25           \\ \hline
$D_{\text{KL}}$  & 4.0877 & 1.4109 & $1.4043$ & $7.2\times 10^{-3}$ & $5.9\times 10^{-3}$ \\ \hline
\end{tabular}
%}
\caption{\textcolor{black}{Kullback–Leibler divergence from the exact distribution to the approximated normal distribution.}}
\label{tab:KL}
\end{table}
\end{remark}
}

\subsection{OMA Baseline}
In OMA, user $t$ and $r$ upload their messages via orthogonal time/frequency resource blocks. In terms of user outage probabilities, the outage probabilities for user $r$ and $t$ in OMA are given as follows:
\begin{align}
    &P^{r/t,\text{O}}_{out} = \mathrm{Pr}\{\mathrm{SNR}^{\text{O}}_{r/t} < 2^{2\Tilde{R}_{r/t}}-1 \}
\end{align}
where $\mathrm{SNR}^\text{O}_{r/t} = {p \big|\sum_{m=1}^M g^{r/t}_m  h_m  e^{j\phi^{R/T}_m}\big|^2}/ {\sigma^2_0}$ is the SNR for OMA. The closed-form analytical outage probabilities for OMA scenario can be obtained similar to those of NOMA scenario. Due to page limitation, we omit these analytical results and demonstrate them in simulations results.

\section{Numerical Results}\label{num}

\begin{figure}[b!]
    \begin{center}
        \includegraphics[scale=0.5]{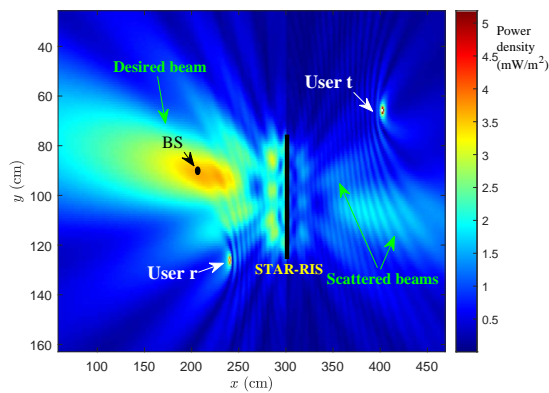}
        \caption{Radiation density of the STAR-RIS-aided uplink transmission.}
        \label{pattern}
    \end{center}
\end{figure}

In this section, simulation results are provided to investigate the performance of the STAR-RIS-aided uplink communication. For our simulations, we assume that the STAR-RIS is a uniform planar array consisting of $M$ elements. The spacing between adjacent elements is half of the carrier wavelength. 
All STAR-RIS-user channels are modeled as Rician fading channels with path loss exponents of $\alpha = 2.2$ and the Rician factor of $K=1.3$ dB.

\begin{figure}[b!]
    \begin{center}
        \includegraphics[scale=0.5]{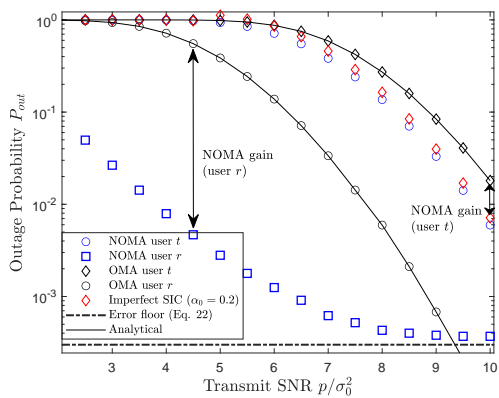}
        \caption{\textcolor{black}{Comparing outage probabilities of OMA and NOMA with perfect/imperfect SIC.}}
        \label{op1}
    \end{center}
\end{figure}

In Fig.~\ref{pattern}, the power density is plotted for the considered uplink transmission where two users are located on different sides of the STAR-RIS where $M = 8\times 8$. The transmission power of the users are set to $p = 10$ dBm. The T\&R phase-shift coefficients are configured according to the cophase condition and the amplitudes are equal, i.e., $\beta^R=\beta^T = 1/\sqrt{2}$. It can be observed that the STAR-RIS significantly improves the signal strength at the location of the BS. However, there are still scattered beams on the other side of the STAR-RIS. These beams consist the signal of user $r$ that is transmitted (leaked) through the STAR-RIS and the signal of user $t$ that is reflected back.

In Fig.~\ref{op1}, the outage probabilities for NOMA and OMA are plotted where the target data rate for both users is set at $\tilde{R}_t = \tilde{R}_r = 1.5$ bit/s/Hz. The noise power for both users is set to $\sigma^2_0 = -50$ dBm and the transmit SNR is varied from $0$ to $10$ dB. Moreover, the amplitudes of the transmission coefficients of the elements are set to $\beta^T=0.2$.
By comparing the user outage probability of NOMA and OMA, it can be observed that the dual-sided STAR-RIS-aided uplink NOMA performs mostly better than STAR-RIS-aided OMA for both users. Thus, NOMA is a good candidate to exploit the \textit{STAR} capability of dual-sided STAR-RISs. In addition, for NOMA with imperfect SIC, the outage probability of user $r$ is unaffected while the outage probability of user $t$ is slightly higher than NOMA with perfect SIC. However, due to the presence of the error floor of uplink NOMA, the performance of NOMA starts to saturate after the transmit SNR is high. The dual-sided STAR-RIS is able to further lower this error floor as we will see in the next figure.

\begin{figure}[b!]
    \begin{center}
        \includegraphics[scale=0.5]{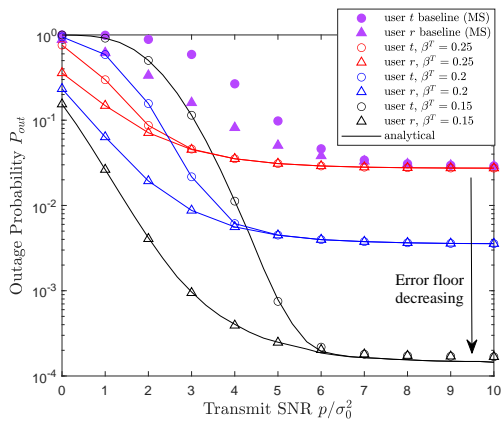}
        \caption{\textcolor{black}{Outage probabilities of NOMA in the high transmit SNR regime.}}
        \label{op2}
    \end{center}
\end{figure}
In Fig.~\ref{op2}, the outage probability error floor is plotted with the simulated outage probabilities for user $t$ and user $r$ employing NOMA.
\textcolor{black}{
In this simulation, we plot the outage probability for the uniform amplitude configuration and for the mode switching (MS) configuration. For the uniform amplitude configuration, the amplitude of the transmission coefficient are set as $0.25$, $0.2$, and $0.15$. For the MS configuration, $25\%$ of the elements are transmitting-only and the remaining elements are reflecting-only.
As can be observed, the MS configuration has the same error floor as the uniform amplitude configuration with $\beta^T = 0.25$. However, the error floor can only be achieved with higher transmit SNR for the MS configuration.
}
Moreover, the error floors for both users can be lowered by adjusting the power ratio between the reflection and transmission for the uniform amplitude configuration.

\section{Conclusions}
In this letter, we provided the hardware model and signal model for the \textit{dual-sided} STAR-RISs.
For the case where the STAR-RIS have surface electric and magnetic impedance, the elements have symmetrical T\&R coefficients.
We derived expressions for the outage probability of a STAR-RIS assisted two-user uplink communication system in high transmit SNR regime. The analytical and numerical results demonstrated the performance gain of NOMA over OMA. It is also revealed that the error floor for the uplink NOMA transmission can be lowered by adjusting the power ratios of STAR elements.

\bibliographystyle{IEEEtran}
\bibliography{mybib}

\end{document}